\documentclass{article}
\usepackage[T1]{fontenc} 
\usepackage[utf8]{inputenc} 
\usepackage{ismir,amsmath,cite}
\usepackage{graphicx}
\usepackage{multirow}
\usepackage{graphicx}
\usepackage[table,xcdraw]{xcolor}
\usepackage[normalem]{ulem}
\useunder{\uline}{\ul}{}

\usepackage[inline,shortlabels]{enumitem}
\setlist[itemize]{noitemsep, topsep=0pt,leftmargin=11pt}
\setlist[enumerate]{noitemsep, topsep=0pt,leftmargin=11pt}
\def\mysize{9pt}
\usepackage[fontsize=9.88]{scrextend}

\usepackage[bookmarks=false,hidelinks]{hyperref}

\title{Optimizing Feature Extraction for Symbolic Music}

\multauthor
{Federico Simonetta$^1$ \hspace{1cm} Ana Llorens$^2$ \hspace{1cm} Martín Serrano$^1$} { \bfseries{Eduardo García-Portugués$^3$ \hspace{1cm} Álvaro Torrente$^{1,2}$}\\
$^1$ ICCMU - Instituto Complutense de Ciencias Musicales, Madrid\\
$^2$ Universidad Complutense de Madrid, Madrid\\
$^3$ Universidad Carlos III, Madrid\\
{\tt\small didone@iccmu.es}
}

\def\authorname{F. Simonetta, A. Llorens, M. Serrano, E. García-Portugués, and Á. Torrente}

\begin{document}

\maketitle

\begin{abstract} 
  
This paper presents a comprehensive investigation of existing feature extraction tools for symbolic music and contrasts their performance to determine the set of features that best characterizes the musical style of a given music score. In this regard, we propose a novel feature extraction tool, named musif, and evaluate its efficacy on various repertoires and file formats, including MIDI, MusicXML, and **kern. Musif approximates existing tools such as jSymbolic and music21 in terms of computational efficiency while attempting to enhance the usability for custom feature development. The proposed tool also enhances classification accuracy when combined with other sets of features. We demonstrate the contribution of each set of features and the computational resources they require. Our findings indicate that the optimal tool for feature extraction is a combination of the best features from each tool rather than those of a single one. To facilitate future research in music information retrieval, we release the source code of the tool and benchmarks.

\end{abstract}

\section{Introduction}\label{sec:introduction}

Feature extraction is a pivotal task in contemporary machine learning. Music features can be categorized into two main types: symbolic and audio. While audio features have been subject to extensive research, computational techniques for symbolic music remain comparatively underexplored.

In recent years, there has been an increasing interest in analyzing symbolic scores in music. This encompasses studies on composer~\cite{kempfert2020where} and style recognition~\cite{herlands2014machine}, affective computing~\cite{qiuNovelMultiTaskLearning2022}, music generation~\cite{zeng2021MusicBERTSymbolicMusic}, analysis of performance~\cite{jeong2017note}, and interpretation~\cite{simonetta2022acousticsspecific}. 
The symbolic dimension of music concerns the conceptual representation of musical data~\cite{vinet2004representation}. This level has been used in the field of Music Information Retrieval (MIR), with particularly successful outcomes when employed to support multimodal approaches~\cite{simonetta2019MultimodalMusicInformation}, which integrate both audio and symbolic levels through audio-to-score alignment techniques~\cite{simonetta2021audiotoscore}.
The symbolic level is also crucial for musicologists, as music scores are the most common source for historical music studies. Musicologists rely on computational tools to extract and analyze musical scores on a large scale~\cite{anuario,choro}. However, traditional manual annotations, such as harmony~\cite{neuwirth2018AnnotatedBeethovenCorpus} and cadence~\cite{hentschel2021AnnotatedMozartSonatas}, are time-consuming and prone to errors. Therefore, computational tools are essential for efficient and accurate musicological analysis.
Presently, two primary tools are available for extracting features from symbolic music: jSymbolic~\cite{mckay2018jSymbolic} and music21~\cite{cuthbert2011feature}. Although both tools are open-source and widely employed, no comprehensive comparison between them has been conducted yet.

In this paper, we propose a novel set of features that is specifically, although not exclusively, tailored for the analysis of 18th-century Italian opera. We have developed a tool for extracting these features, named musif, that is being used for the analysis of operatic music in the Didone project\footnote{\url{https://didone.eu}}~\cite{torrente_musicology_2022}. Here, we conduct a comparative study between musif and other existing tools, thus providing valuable insights into the strengths and weaknesses of each of them. Additionally, we evaluate the efficiency of each tool and demonstrate that musif adds useful features to both music21 and jSymbolic. We observe that, in most cases, a combination of features from multiple tools yields the most powerful feature set. To validate our findings, we test all three tools on various repertoires. We aim to compare the feature sets on file formats with varying levels of representation abilities, such as MIDI, MusicXML, and **kern. While MIDI is widespread in computational studies, it is relatively simplistic for written music; MusicXML and **kern, instead, are less commonly utilized in MIR but provide more accurate representations when dealing with music scores.

The main contributions of this paper are, therefore, threefold. Firstly, we present a new set of features designed for the study of an under-represented repertoire in music computing literature, i.e., 18th-century Italian opera. Secondly, we introduce musif, a new efficient, extensible, and open-source Python tool for feature extraction from symbolic music. Finally, we provide a benchmark of music21, jSymbolic, and musif on a variety of repertoires and file formats.

The whole code used for this study, as well as the code used for the proposed tool, is available at \url{https://github.com/DIDONEproject/music_symbolic_features/}.

\section{Feature Extraction Tools}\label{sec:tools}
\begin{table*}[t]
  \centering
  \caption{Computational efficiency of the three feature extraction tools. Each run was repeated twice and the second run times are indicated between parentheses.}
  \label{tab:efficiency}
  \resizebox{\textwidth}{!}{%
  \begin{tabular}{|c|c|cccccc|}
  \hline
  \textbf{File format}               & \textbf{Tool}      & \multicolumn{1}{c|}{\textbf{Avg CPU Time (s)}} & \multicolumn{1}{c|}{\textbf{Avg Real Time (s)}} & \multicolumn{1}{c|}{\textbf{Avg RAM (GB)}} & \multicolumn{1}{c|}{\textbf{Max RAM (GB)}} & \multicolumn{1}{c|}{\textbf{Tot. errored files}} & \textbf{Tot. files}    \\ \hline
  \multirow{3}{*}{\textit{MIDI}}     & \textit{musif}  & 66.30 (13.30)                                  & 5.62 (\textbf{1.14})                                     & 9.10 (10.1)                                & 14.2 (19.6)                                & 1                                                & \multirow{3}{*}{16734} \\ \cline{2-2}
                                     & \textit{music21}   & 55.3 (55.2)                                    & 4.72 (4.71)                                     & \textbf{7.12 (7.12)}                       & \textbf{9.87} (\textbf{9.94})                       & \textbf{0}                                       &                        \\ \cline{2-2}
                                     & \textit{jSymbolic} & \textbf{2.20 (2.20)}                           & \textbf{1.98} (1.97)                            & 7.97 (7.14)                                & 16.1 (11.7)                                & 14                                               &                        \\ \hline
  \multirow{2}{*}{\textit{MusicXML}} & \textit{musif}  & 15.4 (\textbf{6.63})                                    & 1.32 (\textbf{0.57})                                     & 5.87 (5.12)                                & 12 (10.1)                                  & 4                                                & \multirow{2}{*}{14712} \\ \cline{2-2}
                                     & \textit{music21}   & \textbf{10.8} (10.8)                           & \textbf{0.91} (0.91)                            & \textbf{4.30 (4.33)}                       & \textbf{5.68 (5.55)}                       & \textbf{0}                                       &                        \\ \hline
  \multirow{2}{*}{\textit{**kern}}   & \textit{musif}  & 26 (\textbf{13.1})                             & 2.26 (\textbf{1.14})                            & 5.20 (4.14)                       & 5.60 (4.92)                       & \textbf{0}                                       & \multirow{2}{*}{472}   \\ \cline{2-2}
                                     & \textit{music21}   & \textbf{14.0} (14.1)                           & \textbf{1.21} (1.21)                            & \textbf{3.08 (3.04)}                       & \textbf{4.12 (4.18)}                       & \textbf{0}                                       &                        \\ \hline
  \end{tabular}%
  }
\end{table*}

In this study, we compare three tools for feature extraction from symbolic music: jSymbolic~\cite{mckay2018jSymbolic}, music21~\cite{cuthbert2011feature}, and musif. Other tools such as Humdrum\footnote{\label{footnote:1}\url{https://github.com/humdrum-tools/humdrum-tools}} may be used for feature extraction, but they would require a larger effort for assembling different features from various toolkits and organizing them in a usable tabular format. We will describe each one in detail in the following subsections.

\subsection{jSymbolic}
The jSymbolic tool was initially introduced in 2006~\cite{mckay2006jSymbolic} and subsequently updated in 2018~\cite{mckay2018jSymbolic}. It is an open-source, Java-based software designed to extract features from both MIDI and MEI files. The latest iteration of jSymbolic is capable of extracting 246 distinct features, some of which are multidimensional and account for a total of 1022 values. However, the actual number of extracted features may vary depending on the user's configuration and the musical composition itself. 
jSymbolic features relate to pitch statistics, melodic intervals, chords and vertical intervals, rhythm, instrumentation, texture, and dynamics. In addition to these features, jSymbolic is capable of computing certain characteristics that are not readily available in MIDI files. To achieve this, jSymbolic utilizes the MEI file format to determine the number of slurs and grace notes in a given piece.
While MEI and other high-informative file formats offer additional features such as pitch names, harmonic analysis, and written dynamic or agogic indications, jSymbolic does not take these into consideration.

The jSymbolic software provides users with the flexibility to customize configurations and features, facilitating the integration of previously existing feature values into newer features. Furthermore, users can extract windowed features by specifying window size and overlap in seconds. jSymbolic does not provide pre-built methods for parallel processing of large corpora, thereby requiring the user to implement a suitable strategy. Lastly, jSymbolic provides output options in both CSV and Weka's ARFF format.

The software is accessible as a self-contained program featuring a Graphical User Interface (GUI) and a Command Line Interface (CLI), as well as as a Java library.

\subsection{music21}
music21 is a Python toolkit designed for computational music analysis, which was first introduced in 2010~\cite{cuthbert2010music21}. One of its remarkable features is the capability to parse a wide range of file formats, including MIDI, MusicXML, **kern, ABC, and various others. The music information is represented in an object-oriented hierarchical structure that is aimed at facilitating the development of novel tools.

After its initial academic publication, music21 was further developed with a set of features presented in 2011~\cite{cuthbert2011feature}. The latest version of music21 includes 69 features introduced by jSymbolic, as well as 20 characteristics computed using the information parsed from high-informative file formats. These characteristics are related to key, cadence, harmony, and lyrics.
Regardless of the input file format, music21 consistently outputs 633 features. However, the number of extracted features may vary since some features are zeroed out when they are not computable.

music21 is a Python module that lacks a CLI or a GUI. It does not have a configuration format; rather, it offers a broad range of methods for developing custom pipelines for different types of music information processing. These methods encompass the creation of new features and some automated high-level inference of music characteristics, such as key~\cite{krumhansl1990CognitiveFoundationsMusical}, as well as tools for windowed analysis.

One disadvantage of music21 is that large music scores may result in deeply nested Python objects with numerous non-picklable attributes attached. This makes the programming process challenging, particularly due to the difficulty of saving these objects to a file.

In this study, we have developed a CLI for utilizing music21 feature extraction tools in a manner comparable to musif. This implementation facilitates parallel processing by distributing the extraction of features across numerous files simultaneously.

\subsection{musif}

Our software is named musif~\cite{llorens2023musifa}. It is implemented in Python and built upon the music21 library, and offers an Application Programming Interface (API) with no default settings of significance and a CLI with default settings optimized for most common use cases. 

We leverage music21's internal representation, enabling us to extract features from any file format supported by music21. 
musif is highly customizable and allows users to add custom features as required. After creating the internal representation of the musical score using music21, we extract multiple features and store them in \texttt{pandas} dataframes. This facilitates exporting results in various formats, making musif easily integrable into diverse pipelines.

One limitation of music21 is its restricted ability to serialize complex and large music scores. This restriction also affects the possibility of parallel processing, as Python's single-thread approach necessitates parallelization via processes, which in turn requires context copying and data serialization. Furthermore, parsing large XML files is one of the slowest steps in the feature extraction process. To optimize this procedure, a more favorable strategy would be to store the parsed XML files' logical structure on disk as a cache. We have thus implemented a caching system capable of caching and serializing any music21 object. A restriction to note about the caching system is that the cached scores are read-only. However, this feature enables the writing of parsed scores onto disk and caching of the output from resource-intensive music21 functions into memory.

musif can extract harmony-related features by utilizing standardized harmonic analyses annotated in the MuseScore file format~\cite{neuwirth2018AnnotatedBeethovenCorpus,hentschel2021AnnotatedMozartSonatas}.
Besides, it encompasses a wide range of features, including melodic intervals, harmony, dynamics, tempo, density and texture, lyrics, instrumentation, scoring, and key. Notably, dynamics and tempo are determined by the composer's text notation rather than by MIDI parameters. Furthermore, our implementation includes all features provided by music21 with the exception of 14 features that utilized the caching system in writing mode. The number of extracted features depends on the complexity of the score and is influenced by both the number of parts and musif's compatibility with the encoding. 

NaN values are used to represent non-computable features in a score. For example, when processing datasets with varying instrumentations, some features may not be available for all scores. These values can be replaced with a default value (e.g., 0) or removed from the corpus by deleting either the score or the related feature. In the CLI, we have implemented a heuristic to determine whether a score should be removed from the extracted corpus if it contains too many NaNs. Specifically, we define $r$ as the ratio between the number of columns without NaN and the total number of rows in the output table. If $r<0.1$, we compute $n_i$, which is the number of NaNs in the $i$th row. We remove rows with $n_i$ greater than $\frac{1}{0.99} q_{0.99}$, where $q_{0.99}$ is the $99\%$ quantile of $\{n_1,n_2,\ldots\}$, indicating that 99\% of rows are not deleted. The factor $\frac{1}{0.99}$ can be better understood as dividing the  $Q_{0.99}$ by 99, thus obtaining an estimate of $Q_{0.01}$, and multiplying it by 100, thus obtaining the expected value of $Q_{1.00}$ based on the first $99\%$ of the data. Put differently, it computes the maximum $n_i$ that we expect if the remaining 1\% of rows has a number of NaN ``similar'' to the previous 99\%. Larger values are thus considered outliers. This method was empirically tested on the corpuses used in this work (see Section~\ref{sec:methods}), revealing that only a few scores were generally removed while most lines of the output table were retained. In case a score is not deleted, the CLI removes from the tale the features that are NaN in that score.

musif also incorporates a post-processing module that facilitates the removal, merging, or substitution of values in specific columns or groups of columns within the extracted data. This functionality proves especially advantageous when dealing with large tables generated by musif from a substantial set of scores, as it minimizes the computational effort required for processing such tables.

Like the other tools, we have implemented the capability to extract features at a window level. However, unlike jSymbolic, in our implementation, the window length is specified in musically relevant units such as score measures rather than seconds. This provides more pertinent information for processing music scores.

In contrast to other tools, our solution provides an out-of-the-box capability for processing large corpora through parallel processing, resulting in a reduction of the required time.

The design principles and the features included in musif were presented in a previous publication~\cite {llorens2023musifa}. The code and documentation of musif is available online\footnote{\url{https://github.com/DIDONEproject/musif}, \url{https://musif.didone.eu}}.

\section{Benchmarking Methodology}
\label{sec:methods}

\begin{table*}[t]
  \centering
    \caption{Resulting task size for each dataset and feature set.}
    \label{tab:tasks}
  \resizebox{\textwidth}{!}{%
  \begin{tabular}{|c|c|cccccccc|}
  \hline
                                       &                                    & \multicolumn{1}{c|}{}                                 & \multicolumn{1}{c|}{}                                   & \multicolumn{1}{c|}{}                                   & \multicolumn{5}{c|}{\textbf{Features}}                                                                                                                                                                                        \\ \cline{6-10} 
                                       &                                    & \multicolumn{1}{c|}{}                                 & \multicolumn{1}{c|}{}                                   & \multicolumn{1}{c|}{}                                   & \multicolumn{2}{c|}{\textbf{musif}}                                                 & \multicolumn{2}{c|}{\textbf{music21}}                                                &                                               \\ \cline{6-9}
  \multirow{-3}{*}{\textbf{Extension}} & \multirow{-3}{*}{\textbf{Dataset}} & \multicolumn{1}{c|}{\multirow{-3}{*}{\textbf{Classification task}}} & \multicolumn{1}{c|}{\multirow{-3}{*}{\textbf{Samples}}} & \multicolumn{1}{c|}{\multirow{-3}{*}{\textbf{Classes}}} & \multicolumn{1}{c|}{\textbf{musif}} & \multicolumn{1}{c|}{\textbf{musif native}} & \multicolumn{1}{c|}{\textbf{music21}} & \multicolumn{1}{c|}{\textbf{music21 native}} & \multirow{-2}{*}{\textbf{jSymbolic}}          \\ \hline
                                       & \textit{ASAP performances}         & Composer                                              & 211                                                     & 10                                                      & 710                                    & 91                                            & 633                                   & 602                                          & 225                                           \\ \cline{2-2}
                                       & \textit{ASAP scores}               & Composer                                              & 211                                                     & 7                                                       & 710                                    & 91                                            & 633                                   & 602                                          & 225                                           \\ \cline{2-2}
                                       & \textit{EWLD}                      & Genre                                                 & 2645                                                    & 11                                                      & 710                                    & 91                                            & 633                                   & 602                                          & 225                                           \\ \cline{2-2}
                                       & \textit{JLR}                       & Attribution                                           & 109                                                     & 3                                                       & 732                                    & 113                                           & 633                                   & 602                                          & 226                                           \\ \cline{2-2}
                                       & \textit{Quartets}                  & Composer                                              & 363                                                     & 3                                                       & 1593                                   & 974                                           & 633                                   & 602                                          & 225                                           \\ \cline{2-2}
  \multirow{-6}{*}{\textit{MIDI}}      & \textit{Didone}                  & Decade                                                & 1622                                                    & 8                                                       & 745                                    & 126                                           & 633                                   & 602                                          & 225                                           \\ \cline{1-2}
                                       & \textit{ASAP scores}               & Composer                                              & 211                                                     & 7                                                       & 710                                    & 91                                            & 633                                   & 602                                          & \multicolumn{1}{l|}{\cellcolor[HTML]{C0C0C0}} \\ \cline{2-2}
                                       & \textit{EWLD}                      & Genre                                                 & 3197                                                    & 11                                                      & 724                                    & 105                                           & 633                                   & 602                                          & \multicolumn{1}{l|}{\cellcolor[HTML]{C0C0C0}} \\ \cline{2-2}
                                       & \textit{JLR}                       & Attribution                                           & 109                                                     & 3                                                       & 739                                    & 120                                           & 633                                   & 602                                          & \multicolumn{1}{l|}{\cellcolor[HTML]{C0C0C0}} \\ \cline{2-2}
  \multirow{-4}{*}{\textit{MusicXML}}       & \textit{Didone}                  & Decade                                                & 1636                                                    & 8                                                       & 971                                    & 352                                           & 633                                   & 602                                          & \multicolumn{1}{l|}{\cellcolor[HTML]{C0C0C0}} \\ \cline{1-2}
  \textit{**kern}                      & \textit{Quartets}                  & Composer                                              & 363                                                     & 3                                                       & 734                                    & 115                                           & 633                                   & 602                                          & \multicolumn{1}{l|}{\cellcolor[HTML]{C0C0C0}} \\ \hline
  \end{tabular}%
    }
\end{table*}

To assess the performance of musif in comparison to other tools, we devised a benchmarking methodology. Initially, we identified several datasets that enable testing of diverse file formats. Subsequently, we developed a standardized protocol based on an AutoML pipeline~\cite{feurer2021autosklearn}. We evaluated the computational resources utilized by each tool during extraction and their respective efficacy in various classification tasks.

\subsection{Datasets}

We selected five datasets to evaluate the performance of the tools in analyzing both Standard MIDI Files (SMFs) and highly informative music score formats. For MIDI analysis, we aimed to test both music scores and performances. As for highly informative file formats for music scores, we chose MusicXML and **kern due to their popularity, availability of large datasets, various conversion tools, and compatibility with common music score editing software such as Finale, Sibelius, and MuseScore. While MEI was considered as an option, the limited availability of datasets in this format led us to leave it for future studies.

In this study, we considered the following datasets:
\begin{itemize}
\item \textbf{ASAP}~\cite{foscarin2020asap}: This dataset contains music performances derived from the Maestro dataset~\cite{hawthorne2019enabling} and is synchronized with a corresponding score obtained from the MuseScore's crowd-sourced online library. The dataset comprises 222 music scores in MusicXML and MIDI formats, as well as 1068 music performances in MIDI format. The authors have rectified any significant notation errors found in the music scores. We used this dataset for composer recognition based on music scores and music performances.

\item \textbf{EWLD}~\cite{simonetta2018symbolic}: It contains lead sheets obtained from Wikifonia, a crowd-sourced archive. To reduce errors in music score transcription by inexperienced users, the authors applied algorithmic selection criteria to the dataset. Specifically, they retained only scores with simple notation, without modulations and with a single melodic part. Moreover, all scores contained key signatures and chords throughout. The dataset was augmented by incorporating genre and composer details, as well as the year of first performance, composer birth and death dates, precise title, and additional metadata. This was achieved by cross-referencing the dataset with information sourced from \url{secondhandsong.com} and \url{discogs.com}. We used this dataset for genre recognition.

\item \textbf{Josquin-La Rue}~\cite{cumming2018methodologies}: This dataset was created within the context of the Josquin Research Project and includes 59 Josquin duos and 49 duos by La Rue. The musical scores underwent a meticulous musicological transcription process. Moreover, the music scores were assigned to two labels based on the security of the attribution, thus resulting in four labels (Josquin secure, La Rue secure, Josquin not secure, La Rue not secure). The musical scores are provided in various file formats including MIDI, MusicXML, **kern, Sibelius, and PDF. We used this dataset for composer classification in a real-world attribution problem.

\item \textbf{Quartets}~\cite{sapp2005online}: We retrieved a selection of files from the \url{kern.humdrum.org} website, consisting of all available string quartets in **kern format by Mozart, Haydn, and Beethoven. While the original sources of these musical scores are not always declared, the encoding quality is generally considered to be at a musicological level. In total, we obtained 363 files. We used this dataset for composer classification.

\item \textbf{Didone}~\cite{torrente_musicology_2022}: With the aim of filling an under-studied repertoire, we curated, analyzed, and transcribed over 1600 arias from 18th-century opera, written by dozens of composers. The music scores were transcribed into MusicXML format using Finale Music software and revised by three musicologists independently. Harmonic analyses were added by expert musicologists using MuseScore software in accordance with a prior standard~\cite{neuwirth2018AnnotatedBeethovenCorpus,hentschel2021AnnotatedMozartSonatas} and were reviewed automatically using the ms3 tool\cite{ms3}. We also included various metadata in the database such as year and place of premiere, composer, and high-level formal analysis. This database is an ongoing project and will be made freely available in 2024. We utilized this dataset for classifying the period of composition of each piece, each period being defined in decades (i.e., 1720s, 1730s, 1740s, etc.).

\end{itemize}

\begin{table*}[t]
  \centering
    \caption{Accuracies of AutoML using 10-fold cross-validation on the first ten principal components. The best-performing tool is underlined. The best-performing combination is shown in bold.}
  \label{tab:efficacy-pc}
  \resizebox{\textwidth}{!}{%
  \begin{tabular}{|c|c|c|ccccc|cccc|}
  \hline
                                       &                                    &                                           & \multicolumn{5}{c|}{}                                                                                                      & \multicolumn{4}{c|}{}                                                                                                                                                                                                                                                                                                                                      \\
                                       &                                    &                                           & \multicolumn{5}{c|}{\multirow{-2}{*}{\textbf{Tools}}}                                                                      & \multicolumn{4}{c|}{\multirow{-2}{*}{\textbf{Combinations}}}                                                                                                                                                                                                                                                                                               \\ \cline{4-12} 
  \multirow{-4}{*}{\textbf{Extension}} & \multirow{-4}{*}{\textbf{Dataset}} & \multirow{-4}{*}{\textbf{Dummy guessing}} & \textbf{musif}       & \textbf{\begin{tabular}[c]{@{}c@{}}musif\\native\end{tabular}} & \textbf{music21} & \textbf{\begin{tabular}[c]{@{}c@{}}music21\\native\end{tabular}} & \textbf{jSymbolic}                & \textbf{\begin{tabular}[c]{@{}c@{}}musif native +\\ music21 native\end{tabular}} & \textbf{\begin{tabular}[c]{@{}c@{}}musif native +\\ jSymbolic\end{tabular}} & \textbf{\begin{tabular}[c]{@{}c@{}}music21 native +\\ jSymbolic\end{tabular}} & \textbf{\begin{tabular}[c]{@{}c@{}}musif native + \\ music21 native +\\  jSymbolic\end{tabular}} \\ \hline
                                       & ASAP performances                  & .100                                  & .960                & .715                 & {\ul .978}   & .976                & .916                 & .972                                                                            & .962                                                                       & \textbf{.980}                                                             & .979                                                                                            \\ \cline{2-3}
                                       & ASAP scores                        & .146                                  & .743                & .644                 & {\ul .781}   & .751                & .780                 & .791                                                                            & .819                                                                       & .819                                                                      & \textbf{.857}                                                                                   \\ \cline{2-3}
                                       & EWLD                               & .091                                  & .201                & .157                 & .212         & .204                & {\ul .257}           & .219                                                                            & .245                                                                       & .242                                                                      & \textbf{.259}                                                                                   \\ \cline{2-3}
                                       & JLR                                & .344                                  & .700                & .642                 & {\ul .779}   & .751                & .722                 & .711                                                                            & \textbf{.751}                                                              & .742                                                                      & .741                                                                                            \\ \cline{2-3}
                                       & Quartets                           & .340                                  & .678                & .668                 & .725         & .711                & {\ul .810}           & .768                                                                            & \textbf{.831}                                                              & .791                                                                      & .822                                                                                            \\ \cline{2-3}
  \multirow{-6}{*}{MIDI}               & Didone                          & .125                                  & .359                & .362                 & .403         & .380                & {\ul .443}           & .414                                                                            & .451                                                                       & \textbf{.479}                                                             & .462                                                                                            \\ \hline
                                       & ASAP scores                        & .171                                  & {\ul .773}          & .669                 & .759         & .745                & \cellcolor[HTML]{D9D9D9} & \textbf{.785}                                                                   & \cellcolor[HTML]{D9D9D9}                                                       & \cellcolor[HTML]{D9D9D9}                                                      & \cellcolor[HTML]{D9D9D9}                                                                            \\ \cline{2-3}
                                       & EWLD                               & .091                                  & {\ul .216}          & .185                 & .215         & .201                & \cellcolor[HTML]{D9D9D9} & \textbf{.231}                                                                   & \cellcolor[HTML]{D9D9D9}                                                       & \cellcolor[HTML]{D9D9D9}                                                      & \cellcolor[HTML]{D9D9D9}                                                                            \\ \cline{2-3}
                                       & JLR                                & .334                                  & {\ul \textbf{.793}} & .663                 & .768         & .756                & \cellcolor[HTML]{D9D9D9} & \textbf{.793}                                                                   & \cellcolor[HTML]{D9D9D9}                                                       & \cellcolor[HTML]{D9D9D9}                                                      & \cellcolor[HTML]{D9D9D9}                                                                            \\ \cline{2-3}
  \multirow{-4}{*}{MusicXML}                & Didone                               & .126                                  & .398                & {\ul \textbf{.399}}  & .384         & .374                & \cellcolor[HTML]{D9D9D9} & .392                                                                            & \cellcolor[HTML]{D9D9D9}                                                       & \cellcolor[HTML]{D9D9D9}                                                      & \cellcolor[HTML]{D9D9D9}                                                                            \\ \hline
  **kern                               & Quartets                           & .340                                  & .713                & .711                 & {\ul .767}   & .763                & \cellcolor[HTML]{D9D9D9} & \textbf{.810}                                                                   & \cellcolor[HTML]{D9D9D9}                                                       & \cellcolor[HTML]{D9D9D9}                                                      & \cellcolor[HTML]{D9D9D9}                                                                            \\ \hline
  \end{tabular}%
  }
\end{table*}

\subsection{Experimental setup}
\label{sec:setup}

After selecting the datasets, a standardized protocol was developed for benchmarking the three aforementioned tools. The protocol is based on an AutoML pipeline~\cite{feurer2021autosklearn} and comprises the following steps:
\begin{enumerate}

\item \textbf{Conversion to MIDI}: The datasets were selected and subsequently converted into MIDI format, resulting in two or three file formats for each dataset: MIDI and either MusicXML or **kern. This step aims to evaluate the impact of notational file formats, such as MusicXML or **kern, on classification tasks. Indeed, although MIDI has limited capacity for representing notational aspects of music, it remains uncertain the extent to which  these aspects can  determine the accuracy of machine-learning algorithms for music symbolic analysis. MusicXML files were converted using MuseScore 3, and **kern files were processed with the Humdrum toolkit\footnote{See footnote~\ref{footnote:1}.}.

\item \textbf{Feature extraction}: Features were extracted from MIDI, MusicXML, and **kern files using the methods detailed in Section~\ref{sec:tools} with default settings and without the use of windows, resulting in one array of features for each file. The purpose of this step was to measure the computational cost of the tools. Therefore, all available files in the datasets were used to obtain a larger number of samples and a more accurate estimation of the computational cost, even if they were discarded in later steps. For instance, MIDI scores were already provided in the ASAP dataset; however, we additionally converted them from the MusicXML files. As a result, we extracted features from more files than necessary.
We created a CLI tool in Python for music21 while we utilized the official CLI tools for jSymbolic and musif. Each file format was processed individually, resulting in CSV files for each format. We calculated the average time and RAM usage of each tool. Furthermore, CPU time was collected as a measure of the required time without parallel processing. Lastly, we documented the number of files for which each tool produced errors.

\item \textbf{AutoML}: A state-of-the-art machine learning approach was employed using the Python module \texttt{auto-sklearn}~\cite{feurer2021autosklearn}. The method utilizes Bayesian optimization with surrogate models based on random forests and generates ensembles of models by exploring a vast array of possible architectures. 10-fold cross-validation was used, and the balanced accuracy averaged across the test folds was observed. The best-performing model's result was used for comparison. To initiate the AutoML process, a list of valid files for each dataset was initially defined, discarding those processed in the previous step but unsuitable for validating the classification task. Subsequently, files were selected for which all tools succeeded in extraction, creating comparable datasets for validation. Finally, classes with a number of samples less than twice the number of cross-validation splits were eliminated from each dataset. Consequently, the number of files and categories used in our study differs from the numbers officially provided by each dataset. The classification task performed depended on the dataset, as shown in Table~\ref{tab:tasks}.

\end{enumerate}

We conducted two primary experiments: one utilizing all of the extracted features and another using only the first ten principal components. To achieve this, we standardized the features and applied PCA to obtain the ten first principal components. The rationale for the latter experiment is that a larger feature space typically requires a longer AutoML optimization process and affects the performance of the trained classifiers. As the tools extract varying numbers of features, this experiment enables a principled comparison of the usefulness of the non-redundant information generated by the different tools by homogenizing the number of variables in the AutoML process. In other words, it helps decouple the AutoML optimization capabilities from the number of features.

Due to the overlap between the features extracted with musif and those with music21 with jSymbolic, we also analyzed the concatenation of music21, jSymbolic, and our features. We also observed the performance of musif and music21 when only the native features were used, i.e. when musif was utilized without music21 features and when music21 was run without jSymbolic features. In the following, we denote these feature sets as ``native''. We run each feature extraction and AutoML experiment on a Linux machine with 32 GB of RAM and an i7-8700 CPU, ending the AutoML procedure after 30 minutes. We also experimented with longer AutoML processes and more powerful machines for the first 5 columns of tables~\ref{tab:efficacy-nopc}~and~\ref{tab:efficacy-pc}, but we noticed no significant change in accuracy.

\begin{table*}[t]
  \centering
    \caption{Accuracies of AutoML using 10-fold cross-validation on all the extracted features. The best-performing tool is underlined. The best-performing combination is shown in bold.}
  \label{tab:efficacy-nopc}
  \resizebox{\textwidth}{!}{%
  \begin{tabular}{|c|c|c|ccccc|cccc|}
  \hline
                                       &                                    &                                           & \multicolumn{5}{c|}{}                                                                                                                      & \multicolumn{4}{c|}{}                                                                                                                                                                                                                                                                                                                                      \\
                                       &                                    &                                           & \multicolumn{5}{c|}{\multirow{-2}{*}{\textbf{Tools}}}                                                                                      & \multicolumn{4}{c|}{\multirow{-2}{*}{\textbf{Combinations}}}                                                                                                                                                                                                                                                                                               \\ \cline{4-12} 
  \multirow{-4}{*}{\textbf{Extension}} & \multirow{-4}{*}{\textbf{Dataset}} & \multirow{-4}{*}{\textbf{Dummy guessing}} & \textbf{musif} & \textbf{musif native} & \textbf{music21}        & \textbf{music21 native} & \textbf{jSymbolic}                               & \textbf{\begin{tabular}[c]{@{}c@{}}musif native +\\ music21 native\end{tabular}} & \textbf{\begin{tabular}[c]{@{}c@{}}musif native +\\ jSymbolic\end{tabular}} & \textbf{\begin{tabular}[c]{@{}c@{}}music21 native +\\ jSymbolic\end{tabular}} & \textbf{\begin{tabular}[c]{@{}c@{}}musif native + \\ music21 native +\\  jSymbolic\end{tabular}} \\ \hline
                                       & ASAP performances                  & .100                                  & .983          & .839                 & .983                & .984                & {\ul .985}                          & .983                                                                            & .985                                                                       & \textbf{.990}                                                             & .988                                                                                            \\ \cline{2-3}
                                       & ASAP scores                        & .146                                  & .843          & .626                 & .877                & {\ul .887}          & .886                                & .911                                                                            & .898                                                                       & \textbf{.912}                                                             & .937                                                                                            \\ \cline{2-3}
                                       & EWLD                               & .0912                                  & .224          & .180                 & .249                & .227                & {\ul .248}                          & .236                                                                            & .250                                                                       & .249                                                                      & \textbf{.251}                                                                                   \\ \cline{2-3}
                                       & JLR                                & .344                                  & .746          & .697                 & {\ul \textbf{.806}} & .761                & .747                                & .789                                                                            & .787                                                                       & .751                                                                      & .774                                                                                            \\ \cline{2-3}
                                       & Quartets                           & .340                                  & .828          & .771                 & .843                & .813                & {\ul .901}                          & .843                                                                            & .896                                                                       & .880                                                                      & \textbf{.904}                                                                                   \\ \cline{2-3}
  \multirow{-6}{*}{MIDI}               & Didone                            & .125                                  & .480          & .429                 & .525                & .508                & {\ul .586}                          & .515                                                                            & .572                                                                       & \textbf{.596}                                                             & .557                                                                                            \\ \hline
                                       & ASAP scores                        & .171                                  & .830          & .710                 & {\ul \textbf{.880}} & .841                & \cellcolor[HTML]{D9D9D9}                & .847                                                                            & \cellcolor[HTML]{D9D9D9}                                                       & \cellcolor[HTML]{D9D9D9}                                                      & \cellcolor[HTML]{D9D9D9}                                                                            \\ \cline{2-3}
                                       & EWLD                               & .091                                  & .251          & .200                 & {\ul \textbf{.266}} & .253                & \cellcolor[HTML]{D9D9D9}                & .245                                                                            & \cellcolor[HTML]{D9D9D9}                                                       & \cellcolor[HTML]{D9D9D9}                                                      & \cellcolor[HTML]{D9D9D9}                                                                            \\ \cline{2-3}
                                       & JLR                                & .334                                  & .797          & .704                 & {\ul \textbf{.815}} & .806                & \cellcolor[HTML]{D9D9D9}                & .750                                                                            & \cellcolor[HTML]{D9D9D9}                                                       & \cellcolor[HTML]{D9D9D9}                                                      & \cellcolor[HTML]{D9D9D9}                                                                            \\ \cline{2-3}
  \multirow{-4}{*}{MusicXML}                & Didone                           & .126                                  & .510          & .504                 & {\ul .527}          & .516                & \cellcolor[HTML]{D9D9D9}{\ul \textbf{}} & \textbf{.535}                                                                   & \cellcolor[HTML]{D9D9D9}                                                       & \cellcolor[HTML]{D9D9D9}                                                      & \cellcolor[HTML]{D9D9D9}                                                                            \\ \hline
  **kern                               & Quartets                           & .340                                  & .822          & .786                 & {\ul .830}          & .820                & \cellcolor[HTML]{D9D9D9}{\ul \textbf{}} & \textbf{.842}                                                                   & \cellcolor[HTML]{D9D9D9}                                                       & \cellcolor[HTML]{D9D9D9}                                                      & \cellcolor[HTML]{D9D9D9}                                                                            \\ \hline
  \end{tabular}%
  }
\end{table*}

\section{Results}
\label{sec:results}

Table~\ref{tab:efficiency} summarizes the comparative computational efficiency of the three tools. It is observed that jSymbolic outperforms the other tools when no parallel processing is employed. This can be attributed to the superior performance of Java language, which facilitates faster I/O operations and parsing of byte-level structures such as MIDI files.
musif's caching system significantly reduces the time required for feature extraction during multiple runs, such as those performed during the development and debugging of newly added features. For MIDI files, the extraction process can be accelerated by a factor of five. When comparing the time needed for extraction, jSymbolic is still faster than musif. However, our caching system is advantageous when a cache is available.
Regarding MusicXML and **kern files, musif and music21 use the same parser engine, making their time values more comparable. In this case, music21 is slightly faster than musif but also attempts to extract a smaller number of features. Nevertheless, musif's the caching system allows for a 50\% reduction in extraction times.
The music21 tool proves to be the optimal choice when taking into account RAM utilization.

Table~\ref{tab:tasks} presents the dataset sizes used in our experiments, which are obtained through the protocol detailed in Section~\ref{sec:setup}. The sample sizes vary from 109 to 3197, while the number of classes ranges from 3 to 11, depending on the dataset. The music21 feature extraction process produces a fixed set of 602 native features, supplemented by an additional 31 features re-implemented from the jSymbolic feature set. In contrast, jSymbolic consistently extracts a set of 225 features with minor variations. musif extracts a variable number of features depending on its ability to parse different music structures, ranging from 91 to 974 extracted features. The remaining features extracted by musif are computed using the music21 feature extraction methods. It is worth noting that music21 always converts non-computable features to zero, whereas musif allows users to assign different values or perform other operations.

\begin{table*}[t]
\centering
\caption{Accuracies of AutoML. Effect of harmonic features on the Didone dataset.}
\label{tab:harmonic}
\resizebox{\textwidth}{!}{%
\begin{tabular}{c|cccc|ccc|}
\cline{2-8}
\multicolumn{1}{l|}{}                                         & \textbf{Extension}         & \textbf{Harmonic features} & \textbf{musif} & \textbf{musif native} & \textbf{musif native + music21 native} & \textbf{musif native + jSymbolic}         & \textbf{musif native + music21 native + jSymbolic} \\ \hline
\multicolumn{1}{|c|}{}                                        &                            & No                         & .359          & .362                 & .414                                  & .451                                     & .462                                              \\
\multicolumn{1}{|c|}{}                                        & \multirow{-2}{*}{MIDI}     & Yes                        & \textbf{.380} & .372                 & .398                                  & .452                                     & \textbf{.465}                                     \\ \cline{2-8} 
\multicolumn{1}{|c|}{}                                        &                            & No                         & .398          & .399                 & .392                                  & \multicolumn{1}{l}{\cellcolor[HTML]{D9D9D9}} & \multicolumn{1}{l|}{\cellcolor[HTML]{D9D9D9}}         \\
\multicolumn{1}{|c|}{\multirow{-4}{*}{\textbf{First 10 PCs}}} & \multirow{-2}{*}{MusicXML} & Yes                        & .385          & \textbf{.406}        & \textbf{.409}                         & \multicolumn{1}{l}{\cellcolor[HTML]{D9D9D9}} & \multicolumn{1}{l|}{\cellcolor[HTML]{D9D9D9}}         \\ \hline
\multicolumn{1}{|c|}{}                                        &                            & No                         & \textbf{.510} & .504                 & .515                                  & \textbf{.596}                            & .557                                              \\
\multicolumn{1}{|c|}{}                                        & \multirow{-2}{*}{MIDI}     & Yes                        & .507          & .437                 & .518                                  & .575                                     & .560                                              \\ \cline{2-8} 
\multicolumn{1}{|c|}{}                                        &                            & No                         & .480          & .429                 & .535                                  & \multicolumn{1}{l}{\cellcolor[HTML]{D9D9D9}} & \multicolumn{1}{l|}{\cellcolor[HTML]{D9D9D9}}         \\
\multicolumn{1}{|c|}{\multirow{-4}{*}{\textbf{All features}}} & \multirow{-2}{*}{MusicXML} & Yes                        & \textbf{.535} & .521                 & \textbf{.564}                         & \multicolumn{1}{l}{\cellcolor[HTML]{D9D9D9}} & \multicolumn{1}{l|}{\cellcolor[HTML]{D9D9D9}}         \\ \hline
\end{tabular}%
}
\end{table*}

Tables~\ref{tab:efficacy-pc} and~\ref{tab:efficacy-nopc} demonstrate the effectiveness of feature sets in representing significant aspects of music analysis across various repertoires. The results in Table~\ref{tab:efficacy-nopc} must be interpreted with caution due to the longer AutoML process required by accurate models when using a higher number of features. Overall, music21 and jSymbolic are effective tools for extracting features from MIDI files, while musif shows promising results for MusicXML files, particularly when utilizing the first ten principal components during validation. This difference in performance can be attributed to the presence of highly correlated features in musif, a consequence of its granularity. We also evaluated combinations of feature sets and found that optimal performance is achieved by employing multiple tools. For MIDI files, jSymbolic is fundamental in achieving model accuracy, but incorporating musif and music21 generally enhances performance. For MusicXML and **kern files, leveraging both musif and music21 yields optimal results, especially when considering the first ten principal components.

When comparing the efficacy of models trained on MusicXML, **kern, and MIDI files, no discernible pattern emerges indicating the superiority of highly informative file formats over SMFs for representing music scores. In fact, the only instances where the MusicXML files exhibit superior performance are in the Josquin-La Rue dataset and genre recognition on the EWLD dataset when all features are utilized. However, for all the remaining tasks, MIDI files demonstrate superior performance. This is likely due to the fact that jSymbolic can only extract features from MIDI files and is simultaneously the most important source of features for music score analysis. Consequently, in this study, the MusicXML and **kern datasets lack some relevant features that can be extracted only when converted to MIDI. Even when comparing only the proposed tool and music21's performances, MusicXML and **kern files do not show a clear advantage over MIDI files, particularly when considering the combination of both tools. It should be noted that jSymbolic can extract features from MEI as well, thus potentially allowing for better performances.

The effect of missing values on tool performance is a significant concern and may be a contributing factor to the comparatively lower results for MusicXML and **kern files. While music21 substitutes all missing values with 0, musif utilizes a hybrid strategy that entails either removing a row or column from the table (refer to Section~\ref{sec:tools}). The most effective method for handling missing values remains an open issue.

We assessed the impact of harmonic features on the Didone dataset using musif. Unfortunately, due to the time-consuming nature of harmonic annotations, we were unable to evaluate these features on the other datasets used in this study. We annotated our dataset of more than 1600 opera arias using the standard established in previous works (see Section~\ref{sec:tools}) and extracted melody- and accompaniment-related features with respect to the local key. The extraction of harmonic features resulted in 22 additional features beyond the 126 listed in Table~\ref{tab:tasks} for MIDI files. For MusicXML files, we extracted 265 additional features, raising the total number of extracted features to 617. We observed an overall improvement in classification accuracy when incorporating harmonic features, as demonstrated in Table~\ref{tab:harmonic}. The only instance where performance was degraded by the inclusion of harmonic features was for MIDI files when all the available features were considered (without PCA). We interpret this degradation as an indication that longer processing times are necessary for AutoML when additional, possibly highly correlated features are introduced.

\section{Conclusion}

This paper presents a comprehensive analysis of tools for extracting features from symbolic music. A strict protocol was defined to compare the tools in terms of efficiency and efficacy across various repertoires and file formats. The results indicate that using multiple tools is the most effective approach, with the optimal tool choice depending on the file format and repertoire.

The study emphasizes the importance of using file formats that are accessible by multiple tools. However, it remains open whether highly informative file formats such as MusicXML, **kern, or MEI are relevant for the automatic classification of symbolic scores. The available set of features indicates that, while these formats remain fundamental for certain types of musicological research, they do not seem to entail a significant advantage for machine learning tasks.

The problem of NaN values in extracted features from music scores remains unresolved. Further research is required to explore optimal approaches for replacing, removing, or inferring missing values in music applications.

Additionally, the new musif tool was proposed, which can process various file formats using the music21 parsing engine. The tool also includes a caching mechanism to speed up feature development. Moreover, motivated by the experiments presented in this work, we included the whole music21 and jSymbolic tools in the newer versions of musif, easing the extraction of the combined feature sets from large corpora.

\section{Acknowledgements}
This publication is a result of the Didone Project, which has received funding from the European Research Council (ERC) under the European Union’s Horizon 2020 research and innovation program, Grant agreement No. 788986. It has also been conducted with funding from Spain’s Ministry of Science and Innovation ({IJC2020-043969-I/AEI/10.13039/501100011033}).

Part of the computational experiments were run at the FinisTerrae III cluster of the Galician Supercomputing Center (CESGA). The authors gratefully acknowledge the access to these resources.

\bibliography{zotero.bib,bibliography.bib}

\end{document}